\documentstyle[titlepage,12pt]{article}
\def\VEV#1{\left\langle #1\right\rangle}
\def\Tilde#1{\widetilde {#1}}
\long\def\@makefntext#1{\parindent 0cm\noindent
\hbox to 1em{\hss$^{\@thefnmark}$}#1}

\begin{document}
\newfont{\bg}{cmr10 scaled\magstep3}
%
\newcommand{\gsimeq}
{\hbox{ \raise3pt\hbox to 0pt{$>$}\raise-3pt\hbox{$\sim$} }}
\newcommand{\lsimeq}
{\hbox{ \raise3pt\hbox to 0pt{$<$}\raise-3pt\hbox{$\sim$} }}


\newcommand{\pletb}[3]{Phys. Lett. {\bf B#1}, (#2) #3}
\newcommand{\prevlet}[3]{Phys. Rev. Lett. {\bf #1}, (#2) #3}
\newcommand{\prevd}[3]{Phys. Rev. {\bf D#1}, (#2) #3}
\newcommand{\nuclpb}[3]{Nucl. Phys. {\bf B#1}, (#2) #3}
\newcommand{\prog}[3]{Prog. Theor. Phys. {\bf #1}, (#2) #3}
\newcommand{\zeitc}[3]{Z. Phys. {\bf C#1}, (#2) #3}
\newcommand{\chp}{H^{+}}
\newcommand{\mhc}{m_{\chp}}
\newcommand{\mt}{m_{t}}
\newcommand{\mb}{m_{b}}
\newcommand{\mc}{m_{c}}
\newcommand{\ms}{m_{s}}
\newcommand{\mw}{m_{W}}
\newcommand{\mwl}{m_{W_1}}
\newcommand{\mwh}{m_{W_2}}
\newcommand{\df}{\frac{1}{(1-x)^4}}
\newcommand{\dt}{\frac{1}{(1-x)^3}}
\newcommand{\dty}{\frac{1}{(1-y)^3}}
\newcommand{\qt}{Q_t}
\newcommand{\bsg}{b \rightarrow s \gamma}
\newcommand{\bsemil}{b \rightarrow ce\bar{\nu}}
\newcommand{\bbsg}{Br(b \rightarrow s \gamma)}
\newcommand{\bbsemil}{Br(b \rightarrow ce\bar{\nu})}
\newcommand{\gbsg}{\Gamma(b \rightarrow s \gamma)}
\newcommand{\gbsemil}{\Gamma(b \rightarrow ce\bar{\nu})}
\newcommand{\cz}{c_\zeta}
\newcommand{\sz}{s_\zeta}
\newcommand{\cdb}{c_{2\beta}}
\newcommand{\sdb}{s_{2\beta}}

\begin{titlepage}
\rightline{UT-662}
\rightline{BA-93-69}
\rightline{December, 1993}

\vspace*{1.5cm}

\begin{center}

\addtocounter{footnote}{1}

{\large \bf Constraints on left--right symmetric models
\\
\vspace{0.5cm}
from the process $\bsg$}%
\\

\vspace{1.3cm}

{\sc K.S. Babu} \\

\vspace{0.4cm}

{\it Bartol Research Institute, University of Delaware \\
Newark, DE 19716, U.S.A.}

\vspace{0.8cm}

{\sc Kazuo Fujikawa} and {\sc Atsushi Yamada} \\

\vspace{.4cm}
{\it Department of Physics,  University of Tokyo \ \
\\       Bunkyo-ku, Tokyo, 113 Japan \\}

\vspace{1.5cm}

{\bf ABSTRACT}
\end{center}

In left-right symmetric models, large contributions
to the decay amplitude $\bsg$ can arise from the mixing of the
$W_L$ and $W_R$ gauge bosons as well as from the charged Higgs boson.
These amplitudes are enhanced by the factor $\mt/\mb$ compared to
the contributions in the standard model.  We use the recent
CLEO results on the radiative $B$ decay to place constraints on the
$W_L-W_R$ mixing angle $\zeta$ and the mass of the charged Higgs boson
$m_{H^\pm}$.  Significant departures from the standard model predictions
occur when $|\zeta| \stackrel{_>}{_\sim} 0.003$ and/or when
$m_{H^\pm} \stackrel{_<}{_\sim}$ a few TeV.

\end{titlepage}
\baselineskip = 0.7cm

\section*{Introduction:}

Left--right symmetric theories of the weak interactions
based on the gauge group $SU(2)_L \times
SU(2)_R \times U(1)_Y$ are attractive extensions of
the standard model possessing manifest parity invariance \cite{pati}.
These theories also have greater quark--lepton symmetry than the standard
model since they require the existence of the right--handed
partner of the neutrino $\nu_R$, leading naturally to non--zero
neutrino masses.  The observed
(V-A) nature of the weak interactions is explained by the
spontaneous breaking of parity along with the breaking of
$SU(2)_R$ gauge symmetry at a scale $v_R \gg m_W$.
If the scale $v_R$ of $SU(2)_R$ breaking is
not much above the weak scale, observable deviations
from the predictions of the standard model are possible.
Flavor changing neutral current processes have proven in the past
to be powerful probes of physics beyond the standard model.
For example, in the context of the left--right symmetric models,
the mass of the charged $W_R$ gauge boson
should exceed about $1.6$ TeV, or else it would
contribute to the $K^0-\overline{K}^0$ mass difference at an
unacceptable level \cite{beall}.

In this paper we study the constraints
on the parameters of the left--right symmetric model arising from
the process $b \rightarrow s\gamma$.  Recently the CLEO collaboration
has reported the first observation of the exclusive decay
$B \rightarrow K^*\gamma$ with a branching ratio \cite{cleo}
\begin{equation}
Br(B \rightarrow K^* \gamma) = (4.5 \pm 1.5 \pm 0.9) \times 10^{-5}~.
\label{eqn:bound}
\end{equation}
Eq. (\ref{eqn:bound})
implies both lower and an upper limits on the inclusive decay
$B\rightarrow X_s\gamma$ [e.g. $Br (B \rightarrow
X_s\gamma) < 5.4 \times 10^{-5}$ at 95\% C.L.].  These numbers are in good
agreement with the standard model predictions and as such,
are sensitive to new physics.

In renormalizable gauge theories, the radiative decay
$\bsg$ proceeds through the magnetic moment
operators
$\overline{b}_R\sigma_{\mu \nu} s_L F^{\mu \nu}$ and
$\overline{b}_L\sigma_{\mu \nu} s_R F^{\mu \nu}$, where
$F^{\mu \nu}$ is the electromagnetic field strength tensor.
In the standard model, the $\bsg$ amplitude is proportional to $m_b$ or
$m_s$,
the mass of the bottom quark or the strange quark, because
the pure (V-A) structure of the charged currents requires the
chirality-flip to proceed only through
the mass of the initial or the final
state quark.
In contrast, in left-right symmetric models, the mixing of the
$W_L$ and $W_R$ gauge bosons leads also to (V+A) interactions
between the $W_1$ boson and the quarks, where $W_1$ is the lighter mass
eigenstate formed by $W_L$ and $W_R$. In this case,
the $\bsg$ amplitude can be proportional to the top quark mass $\mt$
rather than $\mb$ or $\ms$ since chirality flip can now occur
with the top quark mass in the intermediate state.
This enhancement of the amplitude gives rise to significant departure
of the decay rate $Br(b \rightarrow s\gamma)$
from the prediction in the standard model,
if the $W_L-W_R$ mixing angle $\zeta$ exceeds about $10^{-3}$.

Left--right symmetric models also predict the existence of
a charged Higgs boson that couples to the quarks.
Its contributions to the $\bsg$ amplitude are also proportional to the
top quark mass, and the experimental result (\ref{eqn:bound})
already probes the charged Higgs mass of a few TeV.
This feature should be compared to the charged Higgs contributions
to the $\bsg$ amplitude in the minimal supersymmetric standard
model (MSSM) which
are proportional to $m_b$ or $m_s$.
The experimental result
(\ref{eqn:bound}) excludes a charged Higgs boson lighter than
a few hundred GeV in this case \cite{berger}.
The enhancement of the charged Higgs boson
contributions in left-right
symmetric models stems from the absence of
natural flavor conservation in the Higgs sector of the
model.  In spite of the absence of flavor conservation in left--right
symmetric models, the interactions of the charged Higgs boson to the
quarks are determined in terms of the quark masses,
Cabibbo-Kobayashi-Maskawa (CKM) mixing angles and the ratio of vacuum
expectation values, just as in the MSSM.

Radiative $b$--decays have been studied in the context of left--right
symmetric models in the past in Refs. \cite{coco,asa}.
The effects of the $W_L-W_R$ mixing on the $\bsg$ amplitude
were studied in Ref. \cite{coco}, but the contributions from the charged
Higgs boson were not examined there. Moreover, our result on the contributions
of the $W_L-W_R$ mixing disagrees with that in Ref. \cite{coco}.
The charged Higgs contributions were analyzed in Ref.
\cite{asa}, but the effects of the $W_L-W_R$ mixing and the leading QCD
corrections were not included. In realistic left-right models,
large contributions
to $\bsg$ amplitude arising from the $W_L-W_R$ mixing and those
from the charged Higgs boson are closely related to each other
and they can be simultaneously sizable.
Here we present a comprehensive analysis of both of these contributions
taking into account their correlations
and clarify the implications of the recent $\bsg$
experiment on the parameters in the left--right symmetric models.

\section*{Left-right symmetric models:}

Left-right
symmetric models of weak interactions are based on the
gauge group $SU(2)_L \times SU(2)_R \times U(1)_{B-L}$.  The quarks
($q$) and leptons ($l$) transform under the gauge group as
\begin{eqnarray}
q_L(2,1,{1 \over 3}) &=& \left(\matrix{u \cr d}\right)_L;~~
q_R(1,2,{1 \over 3}) = \left(\matrix{u \cr d}\right)_R \nonumber \\
l_L(2,1,-1) &=& \left(\matrix{\nu \cr e}\right)_L;~~
l_R(1,2,-1) = \left(\matrix{\nu \cr e}\right)_R
\end{eqnarray}
where generation indices have been suppressed.  The minimal Higgs sector
compatible with the see--saw mechanism for small neutrino masses
consists of the multiplets \cite{rnm} $\Delta_L(3,1,2)$, $\Delta_R(1,3,2)$ and
$\Phi(2,2,0)$ which in component form read as
\begin{eqnarray}
\Delta_{L,R} =
\left(
\begin{array}{cc}
\delta^+/\sqrt{2} & \delta^{++} \\
\delta^0 &  -\delta^+/\sqrt{2}
\end{array}
\right)_{L,R}, \Phi=
\left(
\begin{array}{cc}
\phi^0_1 & \phi^+_2  \\
\phi^-_1  & \phi^0_2
\end{array}
\right)~.
\label{eqn:higgs}
\end{eqnarray}
The field $\Delta_R$ is needed for breaking the gauge symmetry
$SU(2)_L \times SU(2)_R \times U(1)_{B-L}$
down to the gauge symmetry in the
standard model and to give Majorana masses to the
right-handed neutrinos. The field $\Phi$ is required for generating the
quark and lepton
masses. The field $\Delta_L$ is present in the theory to maintain
the discrete parity invariance.
Under parity transformation,
$q_L \rightarrow q_R, l_L \rightarrow l_R, \Delta_L
\rightarrow \Delta_R, \Phi \rightarrow \Phi^{\dagger}$ and
$W_L \rightarrow W_R$.

Spontaneous breaking of the gauge symmetry
$SU(2)_L \times SU(2)_R \times U(1)_{B-L}$
down to $U(1)_{EM}$ is achieved by the
vacuum expectation values (VEV) of the neutral Higgs fields
denoted by
\begin{eqnarray}
\VEV{\Delta_{L,R}} =
\left(
\begin{array}{cc}
0 & 0 \\
v_{L,R} & 0
\end{array}
\right),~~ \VEV{\Phi}=
\left(
\begin{array}{cc}
k & 0 \\
0 & k'
\end{array}
\right)
\label{eqn:vev}~~.
\end{eqnarray}
Among the vacuum expectation values
$k$, $k'$ and $v_{L,R}$, the hierarchy $k,k^\prime \ll v_R$
is needed to preserve the success of the standard (V-A) theory.
In this case, another hierarchy
$v_L \ll k,k^\prime$ follows from a detailed analysis of the Higgs
potential \cite{rnm} which yields the relation
$v_L \sim \gamma k^2/v_R$, where $\gamma$ is
some combination of the Higgs quartic coupling constants.
This is a welcome result
since the analysis of the electroweak $\rho$--parameter leads to
the constraint $v_L \lsimeq 10$ GeV \cite{rev}
and a natural realization of the see--saw
mechanism for small neutrino masses requires $v_L \lsimeq $ a few MeV.
In what
follows, we shall work in the limit $v_L \rightarrow 0$, which is
justified for the above reasons.  The VEVs $k$ or
$k^\prime$ can in general have a phase, but we shall assume
this phase to be small. This is also justified from the detailed analysis
 of the Higgs potential \cite{wolf}.
Small non--zero values of $v_L$ or the relative phase
will not alter our conclusions.

The Yukawa Lagrangian involving the quark fields  is given by
\begin{equation}
{\cal L}_Y = \overline{q}_L h \Phi q_R + \overline{q}_L \tilde{h}
\Tilde{\Phi}q_R + h.c.,
\label{eqn:Yuk}
\end{equation}
where $\Tilde{\Phi} \equiv \tau_2 \Phi^* \tau_2$, $h$ and $\tilde{h}$
are $3 \times 3$ hermitian matrices in generation space.  Eq. (5)
leads to the following mass matrices for the up--type and
down--type quarks:
\begin{equation}
M_u = h k + \Tilde{h} k^\prime~,~~M_d = h k^\prime+\Tilde{h} k~.
\end{equation}

In the charged gauge boson sector, the $W_L^\pm$ and $W_R^\pm$ mix with
their mass--squared matrix given by
\begin{eqnarray}
{\cal M}^2 = {{g^2}\over 2} \left(\matrix{k^2+k^{\prime^2} &
-2 k k^\prime \cr -2kk^\prime & 2v_R^2+k^2+k^{\prime^2}}\right)~~.
\end{eqnarray}
The two mass eigenstates are
\begin{eqnarray}
W_1^\pm &=& c_\zeta W_L^\pm + s_\zeta W_R^\pm ~,\nonumber \\
W_2^\pm &=& -s_\zeta W_L^\pm + c_\zeta W_R^\pm~,
\end{eqnarray}
where $s_\zeta = {\rm sin}\zeta$, $c_\zeta={\rm cos}\zeta$, respectively,
and
\begin{equation}
{\rm tan}2\zeta = {{2kk^\prime}\over {v_R^2}}~.
\label{eqn:tanzeta}
\end{equation}
We have defined $m_{W_1} \le m_{W_2}$ with $m_{W_1} \simeq 80$ GeV.
The mass eigenvalues of $W_1^\pm$ and $W_2^\pm$ are given by
\begin{eqnarray}
\mwl^2 = \frac{g^2}{2}(k^2+k'^2 - 4 kk' \sz\cz + 2 v^2_R \sz^2),
\,\,\,\,\,\,\,
\mwh^2 = \frac{g^2}{2}(k^2+k'^2 + 4 kk' \sz\cz + 2 v^2_R \cz^2).
\label{eqn:mwlh}
\end{eqnarray}

The coupling of the lighter charged $W_1$--boson
to the quarks is given by
\begin{eqnarray}
{\cal L}_{W_1}=\frac{g}{2\sqrt{2}}
\left(
\begin{array}{ccc}
 \bar{u}, & \bar{c}, & \bar{t}
\end{array}
\right) \{
\cz \gamma^\mu (1-\gamma_5)+ \sz \gamma^\mu (1+\gamma_5) \}
{W_1}^+_\mu V
\left(
\begin{array}{c}
d  \\
s  \\
b
\end{array}
\right)
+ h.c.,
\label{eqn:lw1}
\end{eqnarray}
where $V$ is the Cabibbo-Kobayashi-Maskawa (CKM) matrix.  Note that since the
up and down mass matrices are hermitian (as the VEVs $k,k^\prime$ are
taken to be real), the right--handed CKM matrix and the left--handed CKM
matrix are equal ($V_L = V_R = V$), which is reflected in Eq. (11).
The interaction (5) leads to the following
coupling of the corresponding (unphysical) Nambu-Goldstone boson $G_1$
to the quarks,
\begin{eqnarray}
{\cal L}_{G_1}=
\frac{g}{2 \sqrt{2} \mwl}
\left(
\begin{array}{ccc}
 \bar{u}, & \bar{c}, & \bar{t}
\end{array}
\right)
\cz \{  (1-\gamma_5) D_u V - (1+\gamma_5) V D_d \}
{G^+_1}
\left(
\begin{array}{c}
d  \\
s  \\
b
\end{array}
\right)
\nonumber \\
+
\frac{g}{2 \sqrt{2} \mwl}
\left(
\begin{array}{ccc}
 \bar{u}, & \bar{c}, & \bar{t}
\end{array}
\right)
\sz \{ (1+\gamma_5) D_u V - (1-\gamma_5) V D_d \}
{G^+_1}
\left(
\begin{array}{c}
d  \\
s  \\
b
\end{array}
\right)
+ h.c.,
\label{eqn:lg1}
\end{eqnarray}
where we define the diagonal mass matrices
$D_u=diag(m_u, m_c, \mt)$ and $D_d=diag(m_d, m_s, \mb)$.
The couplings of the heavier $W_2^\pm$ gauge boson and the
analogous Nambu--Goldstone boson $G_2$ can be obtained from
Eqs. (11)-(12)
by the replacement $s_\zeta \rightarrow c_\zeta, c_\zeta \rightarrow
-s_\zeta$ and $\gamma_5 \rightarrow -\gamma_5$, but are not necessary for
the present analysis.

The model has two physical charged Higgs bosons.  In the limit of
$v_L \rightarrow 0$, one of them, $\delta_L^+$, becomes mass eigenstate
by itself.  The Higgs boson $\delta_L^+$ has couplings only to the leptons,
and does not enter into the discussion of the $\bsg$ amplitude.
The second physical Higgs boson,
$H^\pm$, which is the linear combination orthogonal to $G_{1,2}^\pm$,
\begin{eqnarray}
H^\pm = N_{H^+}\left[k^\prime \phi_1^\pm + k \phi_2^\pm +
{{(k^2-k^{\prime^2})}\over {\sqrt{2}v_R}} \delta_R^\pm\right],
\end{eqnarray}
has the following Yukawa coupling to the quarks:
\begin{eqnarray}
& & {\cal L}_{\chp}=
-\frac{\sin(2\beta)N_{\chp}}{2\cos(2\beta)}
\left(
\begin{array}{ccc}
 \bar{u}, & \bar{c}, & \bar{t}
\end{array}
\right)
\{ (1-\gamma_5) D_u V - (1+\gamma_5) V D_d \}
\chp
\left(
\begin{array}{c}
d  \\
s  \\
b
\end{array}
\right)
\nonumber \\
& &
-\frac{N_{\chp}}{2\cos(2\beta)}
\left(
\begin{array}{ccc}
 \bar{u}, & \bar{c}, & \bar{t}
\end{array}
\right) \{
 (1+\gamma_5) D_u V - (1-\gamma_5) V D_d \}
\chp
\left(
\begin{array}{c}
d  \\
s  \\
b
\end{array}
\right)
+ h.c.,
\label{eqn:lch}
\end{eqnarray}
where
\begin{eqnarray}
N_{\chp}=1/\sqrt{k^2+k'^2 +\frac{(k^2-k'^2)^2}{2v^2_R}},\,\,\,\,\,
\tan\beta=k/k'.
\end{eqnarray}
The mass of this charged Higgs boson is dependent on the detailed structure
of the Higgs potential, and we leave it as a free parameter in our analysis.


\section*{The process {\boldmath $\bsg$}:}

We now investigate the effective
Hamiltonian describing $\bsg$ in the left-right symmetric model.
Using the Lagrangians (\ref{eqn:lw1}), (\ref{eqn:lg1}) and
(\ref{eqn:lch}), the effective Hamiltonian for $\bsg$ decay
can be written as
\begin{equation}
H_{eff} = \frac{e}{16\pi^2} \frac{2 G_F}{\sqrt{2}} V_{tb} V^*_{ts}
m_b (A_L \bar{s}_L \sigma^{\mu\nu} b_R + A_R\bar{s}_R \sigma^{\mu\nu} b_L)
F_{\mu\nu}, \label{eqn:heff}
\end{equation}
with
\begin{eqnarray}
A_L&=&A_{SM}(x) + \zeta \frac{\mt}{\mb} A_{RH}(x) +
\frac{\mt\sdb}{\mb\cdb^2} A^1_{\chp}(y) + \tan^2(2\beta) A^2_{\chp}(y),
\label{eqn:al} \\
A_R&=&\zeta \frac{\mt}{\mb} A_{RH}(x) +
\frac{\mt\sdb}{\mb\cdb^2} A^1_{\chp}(y) + \frac{1}{\cdb^2} A^2_{\chp}(y)
{}~.
\label{eqn:ar}
\end{eqnarray}
Here the masses of the light quarks $u$, $d$, $s$ and $c$ have been neglected
and the approximations $\cz \simeq 1$, $\sz \simeq \zeta$ and $N_{\chp}
\simeq g/(\sqrt{2}\mw) $ have been used.
In eq. (\ref{eqn:heff}), $F_{\mu\nu}$ is the electromagnetic field
strength tensor, $x=\mt^2/\mw^2$, $y=\mt^2/\mhc^2$, $s_{2\beta} =
{\rm sin}2\beta$ and $c_{2\beta}={\rm cos}2\beta$.
The functions $A_{SM}$, $A_{RH}$, $A^1_{\chp}(y)$ and $A^2_{\chp}(y)$
are found to be
\begin{eqnarray}
A_{SM}(x) &=&
\df \qt \{
\frac{x^4}{4} - \frac{3}{2}x^3 +
\frac{3}{4}x^2 + \frac{x}{2} + \frac{3}{2}x^2 \log(x)
\}
\label{eqn:asm} \nonumber\\
&+&
\df \{ \frac{x^4}{2} + \frac{3}{4}x^3 -
\frac{3}{2}x^2 + \frac{x}{4} - \frac{3}{2}x^3 \log(x)
\}, \nonumber\\
A_{RH}(x) &=&
\dt \qt \{
- \frac{x^3}{2} - \frac{3}{2}x + 2 + 3x\log(x)
\}
 \nonumber\\
&+&
\dt \{ -
\frac{x^3}{2} + 6x^2 - \frac{15}{2}x + 2 - 3x^2 \log(x)
\}.
\label{eqn:arh}\nonumber\\
A^1_{\chp}(y) &=&
\dty \qt \{
- \frac{y^3}{2} +2y^2 - \frac{3}{2}y - y\log(y)
\}
+
\dty \{ -
\frac{y^3}{2} + \frac{y}{2} + y^2 \log(y)
\}, \label{eqn:ach}\nonumber \\
A^2_{\chp}(y) &=& {1 \over 3}A_{SM}(y)-A^1_{\chp}(y),
\end{eqnarray}
where $Q_t=2/3$ is the electric charge of the top--quark.

$A_{SM}$ in eq. (\ref{eqn:al}) is the contribution given by the standard
model \cite{inamilim}. The right-handed coupling
in eq. (\ref{eqn:lw1}) leads to the contributions
$(\mt/\mb)\zeta A_{RH}$ \cite{fy} in eqs. (\ref{eqn:al}) and
(\ref{eqn:ar}). These contributions arise from the chirality flip induced by
the top quark mass in the intermediate state, and they are
enhanced by the factor $\mt/\mb$ compared to the
standard model.
Chirality-flip inside the loop is forbidden in the standard model
because of the purely left-handed nature of the W-boson coupling to the quarks.
The last two terms in eqs. (\ref{eqn:al}) and (\ref{eqn:ar})
are the contributions from the charged Higgs boson $\chp$.
Note that these are also enhanced by the
factor $\mt/\mb$ compared to the contributions in the standard model.
This is in contrast with the charged Higgs contributions
in the minimal supersymmetric standard model (MSSM),
which are proportional to $\mb$
or $\ms$, without any large enhancement factor \cite{berger}.
It is worthwhile to emphasize the correlation between the $W_L-W_R$
mixing contributions and the charged Higgs contributions
in eqs. (\ref{eqn:al}) and (\ref{eqn:ar}).
{}From the expressions (\ref{eqn:tanzeta}) and (\ref{eqn:mwlh}),
the mixing angle $\zeta$ can be written as
$\zeta \simeq \sin(2\beta)\cdot(m_{W_1}^2/m_{W_2}^2)$.
Therefore the large contributions coming from the $W_L-W_R$ mixing
and the charged Higgs boson are proportional to each other and
they can be sizable if $k$ and $k'$ are of the same order.


The effective Hamiltonian $H_{eff}$ given
in eq. (\ref{eqn:heff}) has been evaluated at the electroweak scale
($\mu \sim m_W$).  To make contact with the $\bsg$ decay, $H_{eff}$
should be evolved down to lower momentum scale ($\mu \sim m_b$)
by the renormalization group analysis.
The leading QCD corrections to the Hamiltonian (\ref{eqn:heff})
during its evolution
turn out to be significant \cite{fran,qcd}.
These QCD corrections have been computed
in the standard model in Refs. \cite{gsw,ita} by analyzing the operator mixing
between the magnetic moment operator
$\bar{s}_L \sigma^{\mu\nu} b_R F_{\mu\nu}$
in eq. (\ref{eqn:heff})
and the four Fermi operators involving the
quarks lighter than the $W$ bosons.
In left-right models, because of the $W_L-W_R$ mixing,
there exists some new four-Fermi operators which mix with the
magnetic moment operator $\bar{s}_L \sigma^{\mu\nu} b_R F_{\mu\nu}$.
However, the effects of these
new operators are simply order $\zeta$ {\it without
the enhancement factor of} $\mt/\mb$, and are negligible in
our analysis.
The running of the strong coupling constants and the effects of the
operator mixing are also
negligible in the momentum region above
the $W_1$ boson mass,
because of the asymptotic freedom of the strong interactions,
and consequently the effects of new scales characterized by
$\mt$ and $\mhc$ are ignored.
The QCD corrections to the other magnetic moment operator
$\bar{s}_R \sigma^{\mu\nu} b_L F_{\mu\nu}$ can be computed in analogy to the
case of the operator
$\bar{s}_L \sigma^{\mu\nu} b_R F_{\mu\nu}$ because the strong interactions
respect parity.
Therefore, we compute the QCD corrections
to the Hamiltonian (\ref{eqn:heff})
following the procedure of Ref. \cite{gsw,ita} established in the
standard model.
Here we use the simplified analytical results which are exact to within a few
percent \cite{gsw}.
Including the QCD corrections, $A_L$ and $A_R$ in the
effective Hamiltonian (\ref{eqn:heff}) are renormalized as
\begin{eqnarray}
A^{eff}_L &=& \eta^{-32/23}
\{ A_L + \frac{3}{10}X(\eta^{10/23}-1) + \frac{3}{28}X(\eta^{28/23}-1)\},
\nonumber \\
\,\,\,A^{eff}_R &=& \eta^{-32/23}A_R. \label{eqn:alreff}
\end{eqnarray}
with $X=208/81$ \cite{ita}
and $\eta=\alpha_s(\mb^2)/\alpha_s(\mw^2) \simeq 1.8$.
The last two terms in $A^{eff}_L$ come from the operator mixing in the
standard model. Analogous contributions to $A^{eff}_R$ in the standard
model are proportional to the mass of the strange quarks and are
neglected in our analysis.

The branching fraction $\bbsg$ is computed following the procedure
of Ref. \cite{fran} by normalizing the
decay width $\gbsg$ to the semileptonic decay width $\gbsemil$,
\begin{eqnarray}
 \bbsg = \frac{\gbsg}{\gbsemil} \bbsemil,
\label{eqn:bbsg}
\end{eqnarray}
and using $\bbsemil \simeq 11 \%$ \cite{pd}.
Using the Hamiltonian (\ref{eqn:heff}) with renormalized
quantities $A^{eff}_{L,R}$,
the rate for $\gbsg$ normalized
to the semileptonic rate is given by,
\begin{eqnarray}
& &\frac{\gbsg}{\gbsemil} = \frac{3\alpha}{2\pi\rho(\mc/\mb)(1-\delta_{QCD})}
( |A^{eff}_L|^2 + |A^{eff}_R|^2 )
\label{eqn:rate}
\end{eqnarray}
In eq. (\ref{eqn:rate}), $\rho(\mc/\mb)$ and $\delta_{QCD}$ are the
phase space
suppression factor and the QCD corrections to the semileptonic decay,
respectively. These factors are evaluated as $\rho(\mc/\mb)=0.447$ and
$\delta_{QCD} = (2 \alpha_s(\mb^2)/3\pi)f(\mc/\mb)$,
where $f(\mc/\mb) = 2.41$ \cite{lpqcd}. (We take $\alpha_s(\mb^2) =0.23$.)


We investigate implications of eqs. (\ref{eqn:bbsg}) and
(\ref{eqn:rate}) numerically.  In
Fig. 1, we plot the branching ratio as a function of the left--right
mixing angle $\zeta$ for three different values of the top--quark mass,
$m_t$=110 GeV, 140 GeV and 170 GeV. The value of $m_b$ appearing in the
enhancement factor $m_t/m_b$ in eqs. (\ref{eqn:al}) and (\ref{eqn:ar})
is the one evaluated at
$\mu \sim m_t$.  We choose it to be 3 GeV corresponding to a pole
mass of $4.8$ GeV.
Here we have kept the charged Higgs
boson mass $m_{H^\pm}$ rather high ($m_{H^\pm}$ = 20 TeV). In this case
the contributions of the charged Higgs boson are negligible.
The CLEO limit on the branching ratio,
$Br(\bsg) \le 5.4 \times 10^{-5}$ implies a limit
$-0.015 \le \zeta \le 0.003$, with the region around
$\zeta = -0.005$ disfavored. This limit should be compared to the
existing bounds on $\zeta$ in left--right models.  A limit
$|\zeta| \le 0.035$ has been inferred from the measurement of the
$\xi$--parameter in polarized $\mu$ decay \cite{pd},
but it assumes the neutrino
to be a Dirac particle, which is not the case in the popular see--saw
mechanism.  A bound $|\zeta| \le 0.004$ has been derived from
non--leptonic $K$ decays using the MIT bag model and
assuming current algebra and PCAC \cite{bag},
but this bound is clouded by traditional strong interaction
uncertainties.
The bound derived here from the $\bsg$ process holds regardless of the
nature of the neutrino and has considerably less strong interaction
uncertainty.

In Fig. 2, we have plotted the branching ratio as a function of
$\sin(2\beta)$ for various Higgs masses ($m_{H^\pm} = 0.5, 1,3,5$ and
$10$ TeV) with $m_t = 140$ GeV kept fixed.
Here we have set $\zeta=0$, so that the entire
non--standard contribution arises from the charged Higgs sector.
When $\sin(2\beta)$ approaches $\pm 1$ (keeping the CKM matrix elements
fixed), the coupling of the charged Higgs boson to the quarks diverges.
By requiring that these couplings should be perturbatively controllable,
i.e., less than $4\pi$, we find that $|s_{2\beta}| \lsimeq 0.98$.
Charged Higgs boson with the mass below a few TeV will
contradict the CLEO results on $\bsg$ if $k$ and $k^\prime$ are of the
same order.

In Fig. 3, we plot contours of branching ratios for various Higgs boson
masses ($m_{H^\pm} = 0.5,1,3,5,10$ TeV) for non--zero values of
$\zeta$ and $m_t=140$ GeV.
We have fixed $m_{W_2} = 1.6$ TeV for this graph (which
satisfies the indirect bound from $K^0-\overline{K}^0$ mass difference)
and used the relation $\zeta \simeq \sin(2\beta)\cdot m_{W_1}^2/m_{W_2}^2$
to determine $\zeta$ for a given $\sin(2\beta)$.
The contributions from the mixing angle $\zeta$ and the charged Higgs
boson {\it act additively} to the effective Hamiltonian
(\ref{eqn:heff}).
Consequently, the charged Higgs boson lighter than a few TeV
are excluded for a wide range of parameter space
even after the inclusion of the effects of $\zeta$.
In particular, the charged Higgs boson with a mass
$\mhc \lsimeq 1$ TeV is allowed only if $k$ and $k'$
differ by an order of magnitude or so.   In Fig. 4 we plot analogous
contours, but corresponding to $m_{W_2}=800$ GeV.  A wider range of
parameters are excluded by $\bsg$ in this case.


\section*{Conclusions:}

We have examined the decay $\bsg$ in left-right symmetric models.
Large contribution to $\bsg$ amplitude
arises from $W_L-W_R$ mixing. The physical charged Higgs boson present
in the minimal Higgs sector of the model also yields significant
contribution to the decay amplitude.
Both these contributions are enhanced by the factor $\mt/\mb$
compared to the standard model or the
minimal supersymmetric standard model.
This enhancement stems from the chirality-flip induced by the
right-handed coupling of the $W_1$-boson to the quarks in the case of the
the $W_L-W_R$ mixing. In the case of the charged Higgs contribution,
the enhancement is closely related to
the absence of flavor conservation in the Higgs sector.
Because of these enhanced contributions, the decay $\bsg$
can serve as a sensitive probe to possible signals of
left-right symmetric models.
The recent CLEO results on the radiative $B$ decay lead to the most
stringent and essentially model--independent
bound on the $W_L-W_R$ mixing angle $\zeta$ in a general class of
left-right models: $-0.015 \le \zeta \le 0.003$.
The mass of the charged Higgs boson, which has not been probed
so far by other experiments, is also stringently
constrained by the $\bsg$ experiments. In particular, the charged Higgs
boson mass lighter than about a few TeV
is excluded for a wide range of parameter space.

\section*{Acknowledgments:}

One of us (K.S.B) thanks the Theory Group at the University of Tokyo
for the warm hospitality extended to him during a visit when this work
was initiated. K.S.B is supported by a grant from
the U.S. Department of Energy.

\vspace*{.5in}

\noindent
{\bf Note added} \\ \noindent
While preparing this manuscript, we received a preprint
by P.~Cho and M.~Misiak (CALT-68-1893, hep-ph 9310332).
They examine the effect of the $W_L-W_R$ mixing on the
$\bsg$ amplitude, however, the charged Higgs contributions are
not considered there.
Our result on the contributions of the $W_L-W_R$ mixing
are in agreement with theirs.
We also found that in the calculation of the QCD corrections
to the effective Hamiltonian (\ref{eqn:heff}), the effects of
the new operators ($O_{9,10}$ in their paper)
on the analysis of the operator mixing
are small due to the reasons we explained in the text.
We would like to thank M.~Misiak for a clarifying discussion on
the QCD corrections.


\section*{Figure Captions}
\renewcommand{\labelenumi}{Fig.~\arabic{enumi}}
\begin{enumerate}
\item
The branching fraction $Br(b \rightarrow s\gamma)$
in the left--right symmetric model as
a function of the $W_L^+-W_R^+$ mixing angle $\zeta$.  The three
curves correspond to $m_t$=110 GeV, 140 GeV, and 170 GeV.  The charged
Higgs boson mass is set equal to 20 TeV
so that its effects are negligible for these curves.

\item
The branching ratio $Br(b \rightarrow s\gamma)$
as a function of the mixing angle
$\sin(2\beta)$ for various values of the charged Higgs boson mass.
$m_{H^+}=$ 0.5 TeV (inner solid), 1 TeV (inner dotdash),
3 TeV (dash), 5 TeV (outer dotdash) and 10 TeV (outer solid).
The top mass is fixed as $m_t$=140 GeV.
This graph corresponds to $\zeta=0$.

\item
$Br(b \rightarrow s\gamma)$ versus $\sin(2\beta)$ for
$m_t$=140 GeV and $m_{W_2}$=1.6 TeV for the same set of $m_{H^+}$ values
as in Fig. 2.

\item
Same as in Fig. 3, but for $m_{W_2}=800$ GeV.

\end{enumerate}

\end{document}